\documentclass[12pt]{article}
\usepackage[dvips]{graphicx}
\usepackage{amsmath}
\usepackage{amsfonts}
\usepackage{naconf}              

\def\pd #1#2{\frac{\partial #1}{\partial #2}}
\newcommand\R{\ensuremath{\mathbb{R}}}
\newenvironment{algorithm}[5]
  {\bigbreak\begin{minipage}{5in}\null\qquad{\bf Algorithm}
    $#2=\operatorname{\mathtt{#1}}(#3)$ \smallskip \\
   \null\qquad\qquad {\it input}: #4 \\
   \null\qquad\qquad {\it output}: #5
   \begin{enumerate}}
  {\end{enumerate}\end{minipage}\bigbreak}
\begin{document}

\renewcommand{\thefootnote}{\fnsymbol{footnote}}
\title{Adaptive Finite Elements and Colliding Black Holes}
\author{Douglas N. Arnold, Arup Mukherjee, and Luc Pouly}
\date{}
\maketitle
\numberbysection	
\footnotetext{To appear in
{\it Numerical Analysis 1997: Proceedings of the 17th Biennial
Conference on Numerical Analysis}, D.F. Griffiths and G. A. Watson, eds.,
Addison Wesley, 1998.}

\abstract{According to the theory of general relativity, the relative
acceleration of masses generates gravitational radiation.  Although
gravitational radiation has not yet been detected, it is believed that
extremely violent cosmic events, such as the collision of black holes,
should generate gravity waves of sufficient amplitude to detect on
earth.  The massive Laser Interferometer Gravitational-wave Observatory,
or LIGO, is now being constructed to detect gravity waves.  Consequently
there is great interest in the computer simulation of black hole
collisions and similar events, based on the numerical solution of the
Einstein field equations.  In this note we introduce the scientific,
mathematical, and computational problems and discuss the development of
a computer code to solve the initial data problem for colliding black
holes, a nonlinear elliptic boundary value problem posed in an unbounded
three dimensional domain which is a key step in solving the full field
equations.  The code is based on finite elements, adaptive meshes, and
a multigrid solution process.  Here we will particularly emphasize
the mathematical and algorithmic issues arising in the generation of
adaptive tetrahedral meshes.}

\section{Introduction}

In Einstein's theory of general relativity space-time is represented
as a four-dimensional semi-Riemannian manifold.  The geodesics of this
manifold are the paths of freely falling particles and gravity is the
manifestation of the curvature of space-time.  A system of nonlinear
partial differential equations, the Einstein field equations, specify
the relationship between the curvature of the space-time manifold and
the mass-energy it contains.  A consequence of the theory is that
when a mass accelerates it gives rise to tiny ripples in the fabric
of space time, called \emph{gravity waves}.  More precisely, if a
mass accelerates in distant space, gravity waves will propogate from
the mass at the speed of light.  They will be detectable as slight
variations in the lengths of objects at the point of observation, these
variations modulating with time and differing according to the direction
in which the lengths are measured.  Consequently gravity waves may
be regarded as signals which transmit information about the dynamics
of distant space-time.  As a window to the universe not confined to
the electromagnetic spectrum gravity waves have immense potential as
conveyors of information about the universe, and the creation of an
effective gravity wave detector could be an event whose significance to
astrophysics is as great as the invention of the optical telescope or
radio telescope.

Because of their tiny amplitude, gravity waves have eluded detection
until now.  Even very massive cosmological events, such as the
spiralling collision of neutron stars or black holes of several solar
masses at a distance from earth of up to about 100 million light years
(on the order of 1,000 Milky Way diameters), will cause only the
slightest distance changes here at earth---on the order of one part in
$10^{22}$.  Although the measurement of such tiny distance changes has
not been feasible heretofore, many physicists believe that technology
has reached a point where gravity wave detection is possible, and
several large projects to construct gravitational wave observatories
are now underway.  The largest of these is the LIGO project (Laser
Interferometer Gravitational-wave Observatory), presently under
construction in Hanford, Washington and Livingston Parish, Louisiana.
Each of the two LIGO installations, like observatories being built
in Italy, Germany, Japan, and Australia, is essentially a Michelson
interferometer, consisting of a long evacuated L-shaped tube.  A laser
beam is split at the vertex of the tube and bounced back and forth many
times between mirrors at the vertex and at the end of each of the legs.
Phase differences can then be detected to measure changes in lengths
of the legs.  In the case of LIGO, each leg will be about 1.25 meters
in diameter and 4 kilometers long, and it will be necessary to measure
distance changes of about $10^{-18}$ meters.  (For comparison, the
diameter of a hydrogen atom is about $10^{-10}$ meters.)

In addition to the immense technological hurdles in the construction of
LIGO (e.g., the very sophisticated optics required, the need for a large
and near perfect vacuum, and the suppression of a variety of sources
of noise), the project also raises an extremely challenging problem of
scientific computation.  Given a detected gravitational waveform, we
must determine the cosmological events that could have given rise to it.
The key step in solving this inverse problem is, as usual, the solution
of the forward problem, which in this case is the numerical solution
of the Einstein field equations.  Because they are expected to be a
major source of gravitational radiation, and also because they entail
some useful simplifications, many researchers are presently focussing
on the case of binary black hole collisions.  The data of the problem
then consists of the initial masses, positions, and linear and angular
momenta of the black holes, and the goal is to obtain the far field
waveforms generated via numerical solution of the Einstein equations.

The full simulation of the spiralling coalesence of two black holes is
being actively pursued, but is, at present not effectively achievable.
In the next section we describe the derivation of an elliptic boundary
value problem called the binary black hole initial data problem, which
is an important part of the full problem.  In Section~3 we describe the
design principles of a code we have written to solve the binary black
hole initial data problem and other elliptic boundary value problems
in three-dimensional space.  The following two sections describe and
validate algorithms used in connection with mesh refinement in the code.
Finally, we present numerical results.  Much of this material
appeared in the thesis of the second author \cite{Mukherjee}.  Proofs
of two of the theorems, which will not be repeated here, can
be found in \cite{ArnoldMukherjeePouly}.

\section{The Einstein Equations and the Initial Data Problem}

In this section we briefly describe the derivation of the elliptic
boundary value problem for the black hole initial data problem.
For details we refer to
\cite{BowenYork,Cook,Kulkarni,KulkarniShepleyYork,YorkPiran}.
We represent space-time, or a portion of it, by a semi-Riemannian
4-manifold $M$ parametrized by coordinates $x_\alpha$, $\alpha=0,1,2,3$.
(Greek indices will in general range from $0$ to $3$.)  Let $\boldsymbol g$
denote the metric tensor with covariant components $g_{\alpha\beta}$ and
contravariant components $g^{\alpha\beta}$ (so these two $4\times4$ symmetric
matrices are inverse to one another).  The Christoffel connections are defined
as
$$
\Gamma^{\gamma}_{\alpha\beta}=g^{\gamma\delta}\left(
\pd{g_{\delta\alpha}}{x_\beta}
+\pd{g_{\delta\beta}}{x_\alpha}
-\pd{g_{\alpha\beta}}{x_\delta}\right)
$$
(where we use the Einstein's convention of summation over repeated indices).
The Ricci curvature tensor is then
$$
R_{\alpha\beta}
:=\pd{\Gamma^{\gamma}_{\alpha\beta}}{x^\gamma}
-\pd{\Gamma^\gamma_{\alpha\gamma}}{x^\beta}
+\Gamma^\gamma_{m\gamma}\Gamma^m_{\alpha\beta}-
\Gamma^\gamma_{m\beta}\Gamma^m_{\alpha\gamma},
$$
and the Einstein tensor may be obtained simply from the Ricci tensor:
$$
G_{\alpha\beta}=R_{\alpha\beta}-\frac12g^{\gamma\delta}R_{\gamma\delta}g_{\alpha\beta}.
$$
Finally, the Einstein equations may be written
$$
G_{\alpha\beta}=8\pi T_{\alpha\beta},
$$
where the stress-energy tensor $\boldsymbol T$ is a given forcing function.
In particular, in the case of a vacuum---which is sufficient for the
case of black holes, since we only compute outside the holes---the
Einstein equations assert the vanishing of the Einstein, and hence Ricci,
tensor.

The Einstein equations form a system of 10 second order quasilinear
partial differential equations for the 10 components $g_{\alpha\beta}$
of the metric tensor in the 4 independent variables $x_\alpha$.
(Expanded out, each equation involves over 1,000 terms!)  However,
the equations are not independent, since the Bianchi identity implies
that $\nabla^\alpha G_{\alpha\beta}=0$, no matter what the metric
$\boldsymbol g$.  Hence the Einstein system really only asserts six
independent equations.  Correspondingly there is a non-uniqueness of
solutions (gauge freedom) which essentially allows for the arbitrary
specification of four of the metric components.

We now describe the Arnowitt--Deser--Misner $3+1$ approach to the
solution of the Einstein equations.  The metric tensor has signature
$(-,+,+,+)$, and we assume that coordinates are chosen so that the
variable $t=x_0$ is timelike, and that the remaining variables $x_i$
(Latin indices range from 1 to 3) are spacelike.  We shall refer to
the variable $t$ as time and to the $x_i$ as spatial variables.  The
manifold $M$ is then foliated by the spacelike hypersurfaces $\Sigma_t$
given by $t=\text{constant}$.  The metric $\boldsymbol g$ induces a
Riemannian metric on each hypersurfaces, which we denote by $\boldsymbol\gamma$.
Using the restrictions of the functions $x_i$, $i=1,2,3$, as coordinates,
the covariant components $\gamma_{ij}$ are equal to $g_{ij}$.
The complete four dimensional metric $\boldsymbol g$ can
then be reconstructed from $\boldsymbol\gamma$ together
with the \emph{lapse} $\beta_i=g_{0i}$ and the \emph{shift}
$\alpha=\sqrt{g^{ij}\beta_i\beta_j-g_{00}}$.  In view of the gauge
freedom, we may determine the lapse and shift in any convenient
way (just how this is to be done is currently an area of intense
investigation).  Next we separate the Einstein equations into two
separate systems of equations.  If we multiply the equations by the
normal to the hypersurface $\Sigma_t$ we obtain the \emph{constraint
equations}, a system of four equations.  The remaining six equations,
which arise by multiplying by the tangential directions, are referred
to as the \emph{evolution equations}.  These names arise because the
constraint equations do not involve second derivatives with respect
to $t$.  Thus they form a purely spatial system of four second order
differential equations in the 12 dependent variables $\gamma_{ij}$ and
$\partial\gamma_{ij}/\partial t$ posed on each of the foliating
manifolds $\Sigma_t$.  More geometrically significant dependent
variables are obtained by using the components of the extrinsic
curvature tensor
$$
K_{ij}=-\frac1{2\alpha}(\pd{\gamma_{ij}}t -\nabla_j\beta_i-\nabla_i\beta_j)
$$
instead of  $\partial\gamma_{ij}/\partial t$, and this is usually
done.  The problem of determining
a solution $\gamma_{ij}$, $K_{ij}$ on $\Sigma_0$ to the constraint
equations is known as the initial data problem.  Once a solution
to the initial data problem is found, the evolution equations
are a system of differential equations, second order in space and
time, which can be used to determine $\gamma_{ij}$ for $t>0$.
The constraint equations may or may not be imposed at positive times.
It can be shown that if they are satisfied at the initial time
and the evolution equations are satisfied exactly, then they are
satisfied at all times.

The initial data problem is highly underdetermined, and there are many
possible solutions.  One of the simplest approaches of physical relevance
to binary black holes procedes from the assumptions
\begin{itemize}
\item The spatial metric is conformally equivalent to a flat metric,
i.e., $\gamma_{ij}=\psi^4\delta_{ij}$ for some
function $\psi:\Sigma_0\to\R$ to be determined.
\item The manifold $\Sigma_0$ is maximally embedded, i.e., 
the trace of $\boldsymbol K$ vanishes.
\end{itemize}
(This is a special case of the method of conformal imaging of
J.~York \cite{York}.)  Under these assumptions three of the four
constraint equations (the \emph{momentum constraints}) are linear
and decoupled from the fourth equation, and solutions to them can be
determined analytically.  The remaining equation (the \emph{Hamiltonian
constraint}) takes the form
\begin{equation}
 \label{eqn:hc}
\Delta\psi + H(x)\psi^{-7}  =0,
\end{equation}
where $H(x)$ depends on the solution to the momentum contraint equations
and encodes the positions, masses, and linear and angular momenta of the
black holes.  To obtain initial data for binary blackhole collisions we
wish to solve this equation on $\R^3\setminus(B_1\cup B_2)$ where the
$B_i$ are disjoint balls (their boundaries are the apparent horizons of
the holes).  The equations are subject to Robin boundary conditions
\begin{equation}
 \label{eqn:bc}
\pd{\psi}{r_i}+\frac1{2r_i}\psi=0
\end{equation}
on the hole boundaries (the apparent horizons of the blackholes), and
the condition $\lim_{|x|\to\infty} \psi(x) = 1$ at infinity (so that
the metric is asymptotic to the flat metric far from the holes).  For
numerical purposes the latter condition is usually replaced by an
artificial boundary condition like
\begin{equation}
 \label{eqn:ic}
\pd{\psi}{r}+\frac1r(\psi -1)=0
\end{equation}
on the boundary of a ball $B_0$ about the origin containing the holes well within its
interior.  (Equation (\ref{eqn:ic}) can be derived by a multipole expansion;
cf.~\cite{Cook}.)

\section{A Code for the Black Hole Initial Data Problem}

We thus wish to solve the semilinear boundary value problem consisting
of the PDE (\ref{eqn:hc}) on $B_0\setminus(B_1\cup B_2)$, together with
the boundary condition (\ref{eqn:bc}) on $\partial B_i$, $i=1$, $2$, and
the artificial boundary condition (\ref{eqn:ic}) on $\partial B_0$.  In
a typical computation the hole radii are of the same general magnitude,
as is the distance of their centers from the origin, while the radius
of the containing sphere is taken to be two to four orders of magnitude
larger.  We have designed a code to solve this problem based on the
following design principles:
\begin{itemize}
\item Because of the nontrivial geometry we use \emph{finite elements}.
\item Since the solution varies significantly over only a small portion of the
domain, in the immediate vicinity of the holes, the mesh is refined
\emph{adaptively}.
\item We reduce the nonlinear systems of equations to linear problems with
\emph{Newton's method}.
\item We solve the resulting large systems of linear equations using \emph{multigrid} techniques
based on the sequence of adaptively generated meshes.
\end{itemize}

The gross structure of a code based on these principles is shown in
Figure~\ref{fig:struct}.  (The requirement of conformity in steps 1 and
2(d) means that every nonempty intersection of two distinct tetrahedra must be
either a common face, a common edge, or a common vertex.)

\begin{figure}[!ht]
\centering
\framebox[5in][c]{\parbox{4.5in}{\raggedright
\begin{enumerate}
\item Begin with an initial coarse conforming tetrahedral mesh matching the
geometry
and some initial approximation of the solution on that mesh.
\item Do until a sufficiently accurate solution is found:
\begin{enumerate}
  \item On the current mesh discretize the problem using piecewise linear
  finite elements.
  \item Do until the approximation on the current mesh is sufficiently close
  to stationary:
  \begin{enumerate}
    \item  Linearize the finite element problem about the current
    approximation on the current mesh.
    \item Solve the linearized problem
    using full multigrid to generate the next approximation: 
    Beginning with the coarse mesh solution, interpolate
    the solution to the next finer mesh and update it by smoothing on the
    finer mesh and residual correction on the coarser mesh.  Continue this
    process recursively to generate the solution of the linear problem
    on the current mesh.
  \end{enumerate}
  \item Assign error indicators to the elements of the current mesh.
  \item Refine the mesh as indicated.  Refine further to restore
  conformity thus obtaining the next finer mesh.
  \item Use the current solution on the old mesh as the initial approximate
  solution on the new mesh.
\end{enumerate}
\end{enumerate}}}
\caption{Gross structure of the code.}
\label{fig:struct}
\end{figure}

The development of a code along these lines involves many issues.  We will
discuss two of them here:
\begin{itemize}
\item  How to refine a tetrahedron, ensuring  that its descendants
are nicely shaped tetrahedra, even after many generations
of refinement?
\item  How to bring a refined mesh into conformity without over-refining?
\end{itemize}

\section{Tetrahedral Refinement}

\begin{figure}
\centering
\mbox{\includegraphics[width=2in]{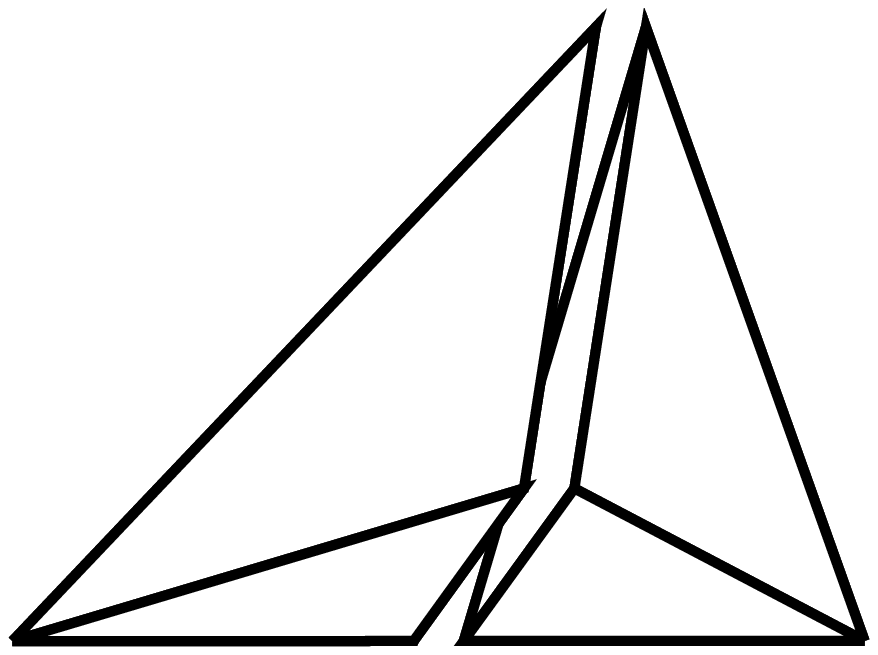}
\includegraphics[width=2in]{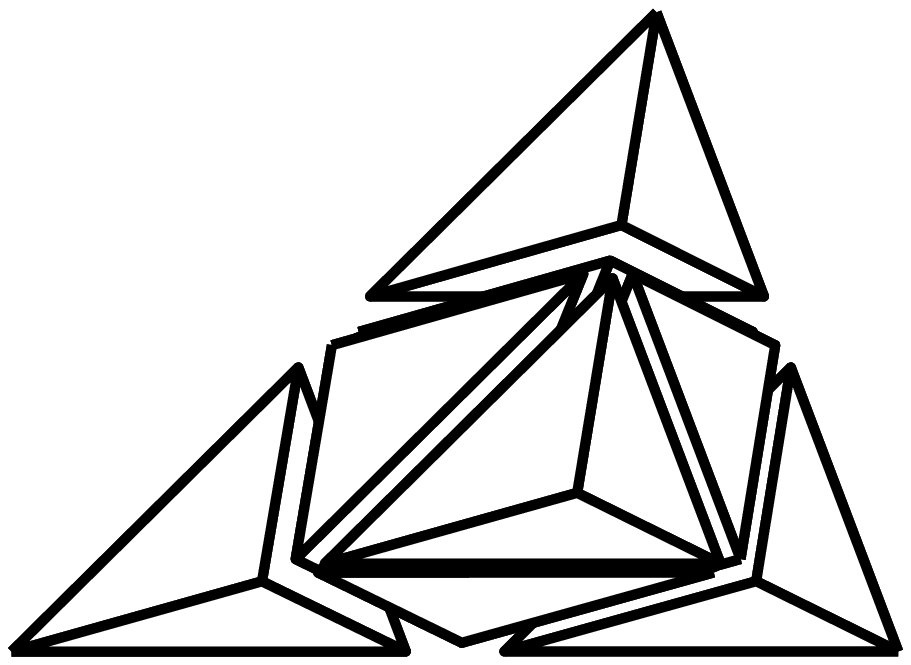}}
\caption{Bisection and octasection.}
\label{fig:bisect}
\end{figure}

The two most natural ways to partition a tetrahedron into subtetrahedra
are bisection (by placing a new vertex on some particular edge, and
connecting it to the existing vertices opposite that edge), and
octasection (cutting off each corner by placing a new face
through the midpoint of the edges emanating from the corner, and then
dividing the remaining octahedron into four tetrahedra).  See
Figure~\ref{fig:bisect}.  The use of octasection requires a great many
intermediate partition strategies if a conforming adaptive mesh is to be
maintained (see \cite{Bey}).  However a conforming adaptive refinement can
be attained using only bisection.  Moreover, since one step of bisection
reduces the element volume by a factor of two rather than eight for
octasection, it can produce element sizes closer to the optimal ones.
Thus we have chosen to base our code exclusively on bisection.

Every time we bisect a tetrahedron we must specify a particular
edge, called the \emph{refinement edge}, on which the new vertex
will be placed.  (We always place the new vertex at the midpoint
of the refinement edge.)  Careful selection of the refinement edge
is essential if the tetrahedra shapes are not to degenerate after
repeated bisections.  For bisection of triangular meshes in two
dimensions, there are two commonly used algorithms for selection of
the bisection edge of a triangle.  One approach is to always select
the longest edge of the triangle.  Rivara \cite{Rivara} has proven
that longest-edge bisection can be applied repeatedly without
degeneration of element shape.  The second approach in two dimensions is
opposite-edge bisection.  In this approach any edge (e.g., the longest)
can be selected for the refinement edge of triangles in the initial
mesh, but as new triangles are created by bisection, they are always
assigned the edge opposite the newly added vertex as the refinement
edge.  For repeated application of opposite-edge bisection there holds
a stronger result than just non-degeneration of shape.  Namely, as is
easily verified, starting with any initial triangle, it and all of its
descendants will belong to at most four distinct similarity classes.  In
three dimensions the selection of a suitable refinement edge is more
complicated.  Longest-edge bisection obviously generalizes to three
dimensions, and has been used successfully \cite{Rivara,RivaraLevin},
but it is unknown whether it avoids shape degeneracy in general.  It is
not at all obvious how to adapt the opposite-edge bisection algorithm to
three dimensions.  We now present such an algorithm and state a theorem
from \cite{ArnoldMukherjeePouly} which shows that, like opposite-edge
bisection in two-dimensions, repeated application of the algorithm
beginning with an arbitrary tetrahedron gives rise to only a fixed
finite number (namely 36) of non-similar element shapes.

Key to our algorithm is a data structure that we call a \emph{marked
tetrahedron}.  Namely not only do we associate to each tetrahedron
a refinement edge, but also to each face a marked edge.  For faces containing
the refinement edge, the marked edge is required to coincide with the
refinement edge.  The marked edges of the other faces will be used
as the refinement edges of the children tetrahedra.  We also associate
to each marked tetrahedron a boolean flag.  If the tetrahedron is \emph{planar},
which means that its marked edges are all coplanar, then the flag may
be either set or unset.  Otherwise it is irrelevant.

Note that when we bisect a tetrahedron, each child inherits one face
from the parent; has one \emph{new face}, interior to the parent; and has two
\emph{cut faces}, strict subsets of a parent face.  With this terminology
we can state our bisection algorithm.

\begin{algorithm}{BisectTet}{\{\tau_1,\tau_2\}}{\tau}
{marked tetrahedron $\tau$}{marked tetrahedra $\tau_1$ and $\tau_2$}
\item Bisect $\tau$ by joining the midpoint of its refinement edge to
each of the two vertices not lying on the refinement edge.
\par \quad Mark the faces of the children as follows:
\item The inherited face
inherits its marked edge from the parent, and this marked edge is the
refinement edge of the child.
\item On the cut faces of the children
mark the edge opposite the new vertex with respect
to the face. 
\item The new face is marked the same way for both
children.  If the parent is planar and flagged, the marked edge is the
edge connecting the new vertex to the new refinement edge.  Otherwise
it is the edge opposite the new vertex.
\item The flag is set in the children if and only if the parent is type
planar and unflagged.
\end{algorithm}

Although quite different in form, and not involving the marked
tetrahedron data structure, several algorithms in the literature can
be shown to produce essentially the same sequence of tetrahedra as
$\mathtt{BisectTet}$ \cite{Baensch, LiuJoe, Maubach}.  Liu and Joe
proved that repeated application of their algorithm gives rise to at
most $168$ similarity classes.  Maubach's algorithm, unlike the others,
applies to bisection of a simplex in any number of dimensions (although
not to a general simplicial mesh).  In the context of Maubach's
algorithm we proved a bound on the number of similarity classes
\cite{ArnoldMukherjeePouly}.

\begin{theorem}
When an $n$-simplex is bisected repeatedly with
this algorithm, there arise at most $2^{n-2}n!$ similarity classes of
each generation and the set of similarity classes depends only on
the generation modulo $n$.
\end{theorem}

Thus in two dimensions there are only two classes of each generation and only
four total.  In three dimensions the corresponding numbers are
12 and 36.  By computation on a particular tetrahedron we showed that
these numbers are sharp \cite{ArnoldMukherjeePouly}.  Maubach recently
proved that the result is sharp for all $n$ \cite{MaubachPreprint}.

\section{Mesh conformity}

In order to implement step 2(d) of the outline given in Figure~\ref{fig:struct},
we need an algorithm that begins with a conforming tetrahedral mesh
and a set of elements selected, and returns a conforming
refinement of the mesh in which all the selected elements have been bisected.
In this section we describe an algorithm based on $\mathtt{BisectTet}$
to accomplish this.

Before stating the algorithm we fix some terminology.  A {\it mesh}
$\mathcal T$ of a domain $\Omega$ in $\mathbb R^3$ is a set of closed
tetrahedra with disjoint interiors and union $\bar \Omega$.  A mesh is
{\it conforming}\/ if the intersection of two distinct tetrahedra is
either a common face, a common edge, a common vertex, or empty.  If
$\nu$ is a vertex of some tetrahedron in the mesh and $\nu$ belongs to
another tetrahedron $\tau$ but is not a vertex of $\tau$, we say that
$\nu$ is a {\it hanging node}\/ of $\tau$.  A mesh is {\it marked}\/
if each tetrahedron in it is marked. A marked conforming mesh is {\it
conformingly-marked}\/ if each face has a unique marked edge (that is,
when a face is shared by two tetrahedra, the marked edge is the same
for both). The tetrahedra of any conforming mesh can be marked so as to
yield a conformingly-marked mesh.  For example, this is accomplished by
the following procedure.  First, strictly order the edges of the mesh
in an arbitrary but fixed manner, e.g., by length with a well-defined
tie-breaking rule.  Then choose the maximal edge of each tetrahedron
as its refinement edge and the maximal edge of each face as its marked
edge.

We now state the main algorithm of this section.

\begin{algorithm}{LocalRefine}{\mathcal T'}{\mathcal T,\mathcal S}
  {conformingly-marked mesh $\mathcal T$ and $\mathcal S \subset \mathcal T$}
  {conformingly-marked mesh $\mathcal T'$}
\item $\bar\mathcal T=\operatorname{\mathtt{BisectTets}}(\mathcal T,\mathcal S)$
\item
$\mathcal T'=\operatorname{\mathtt{RefineToConformity}}(\bar\mathcal T)$
\end{algorithm}

\noindent
The algorithm in the first step, $\mathtt{BisectTets}$, is trivial: we
simply bisect each tetrahedron in $\mathcal S$: 
$$
\operatorname{\mathtt{BisectTets}}(\mathcal T,\mathcal S)=(\mathcal T\setminus\mathcal
S)\cup
\bigcup\limits_{\tau\in\mathcal S}
\operatorname{\mathtt{BisectTet}}(\tau).
$$
In the second step, we perform further refinement
as necessary to obtain a conforming mesh:
\begin{algorithm}{RefineToConformity}{\mathcal T'}{\mathcal T}
{marked mesh $\mathcal T$}{marked mesh $\mathcal T'$ without hanging nodes}
\item set $\mathcal S= \left\{ \tau\in\mathcal T \,|\, \tau \text{ has a hanging
node} \right\}$
\item if $\mathcal S\neq \emptyset$ then
\begin{enumerate}
\item[] $\bar\mathcal T = \operatorname{\mathtt{BisectTets}}(\mathcal T,\mathcal S)$
\item[] $\mathcal T'=\operatorname{\mathtt{RefineToConformity}}(\bar\mathcal T)$
\end{enumerate}
\item else
\begin{enumerate}
\item[] $\mathcal T'=\mathcal T$
\end{enumerate}
\end{algorithm}

\begin{figure}
\centering
\includegraphics[width=2in]{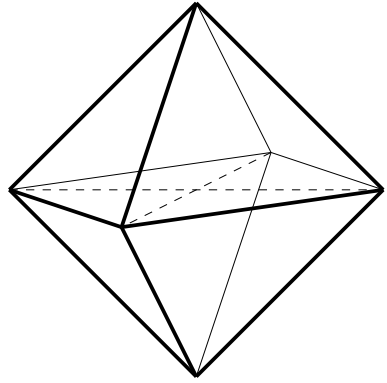}
\caption{A non-conforming mesh without hanging nodes (the barycenter
is {\it not}\/ a vertex of the mesh).}
\label{fig:octahedron}
\end{figure}

\noindent
The recursion in the algorithm $\mathtt{RefineToConformity}$
could conceivably continue forever.  Moreover, even if the
recursion terminates, the output mesh may not be conforming (a
mesh without hanging nodes can nonetheless be non-conforming;
cf., Fig.~\ref{fig:octahedron}).  However, the following theorem, which
is proved in \cite{ArnoldMukherjeePouly}, ensures that the recursion
does terminate in the application of $\mathtt{RefineToConformity}$ in
algorithm $\mathtt{LocalRefine}$ and that the resulting output mesh
is conformingly-marked.  Moreover, it gives a bound on the amount of
refinement which can occur before termination.

\begin{theorem}\label{thm:conform}
Let $\mathcal T_0$ be a conformingly-marked mesh with no
flagged tetrahedra.  For $k=0,1,\ldots$, choose $\mathcal
S_k\subset\mathcal T_k$ arbitrarily, and set $\mathcal
T_{k+1}=\operatorname{\mathtt{LocalRefine}}(\mathcal T_k,\mathcal S_k)$.
Then for each $k$, the application of $\mathtt{RefineToConformity}$ from
within $\mathtt{LocalRefine}$ terminates producing a conformingly-marked
mesh, and each tetrahedron in $\mathcal T_k$ has generation at most
$3k$.  Moreover, if the maximum generation of a tetrahedron in $\mathcal
T_k$ is less than $3m$ for some integer $m$, then the maximum generation
of a tetrahedron in $\mathcal T_{k+1}$ is less than or equal to $3m$.
\end{theorem}

The marked tetrahedron data structure, which was important in the last section
to ensure that the number of element shapes remained bounded, is also essential
here to guarantee conformity.  First, the assignment of a marked edge
ensures that two tetrahedra with a common face are not bisected in an
inconsistent manner.  Second, the flag plays a key role.  If we waive the
requirement that the planar marked tetrahedra in the initial mesh are
unflagged, Theorem~\ref{thm:conform} need no longer be true.

\section{Numerical Results}

We tested the code on a variety of simplified problems with known
solutions before attacking the binary black hole initial data problem.
These results, which can be found in the thesis \cite{Mukherjee}, provided
convincing validation of the code.

As a further validation we considered an initial data problem for
a single black hole admitting a radial solution.  In this case
the differential equation (\ref{eqn:hc}) is solved on the domain
$B_0\setminus B_1$ where $B_1$ is a ball of radius $a$ about the
origin representing the black hole and $B_0$ is a concentric ball of
radius $R\gg a$.  We took $a=\sqrt3/2$ and $R=1028a$.  Robin boundary
conditions like (\ref{eqn:bc}) and (\ref{eqn:ic}) are imposed on the
bounding spheres.  For the function $H$ we use an expression proposed by
Bowen and York \cite{BowenYork},
$$
H = 6 \frac{P^2}{r^4} \left( 1 - \frac{a^2}{r^2} \right)^2.
$$
The parameter $P$ represents the linear momentum of the hole.
For this problem the solution is radial and can be given analytically:
$$
\psi = \left( 1 + \frac{2 E}{r} + \frac{6
a^2}{r^2} + \frac{2 a^2 E}{r^3} + \frac{a^4}{r^4} \right)^{1/4} ,
$$
where $E = ( P^2 + 4 a^2 )^{1/2}$.
The ADM-energy, $E$, and ADM-mass, $M$ are two quantities of physical interest which
are often reported in initial data problem computations.
They are defined as
$$
E = \lim_{R\to\infty}\left[\frac1{16\pi} \int_{B_0\setminus B_1} H \psi^{-7} \,dv + \frac1{4
\pi a} \int_{\partial B_1} \psi \, ds\right],  \quad M = \lim_{R\to\infty}\left ( \frac1{16\pi} \int_{\partial B_1} \psi^4 \, ds
\right)^{1/2} ,
$$
respectively.  In Table~\ref{table:one} we compare our computed results
for $E$ and $M$ to the analytic values for various values of $P$ (all
quantities being scaled by $a$).  In all cases the finest mesh had less
than 70,000 vertices.  Note that, although this problem is essentially
one-dimensional, for validation purposes we ignored this and solved it
as a fully three-dimensional problem.

\begin{table}
\centering
\begin{tabular}{c|ccc|ccc}
      & $E/a$ & $E/a$ & percent & $M/a$ & $M/a$ & percent \vspace{-3pt} \\
\raisebox{7pt}[-2pt]{\mbox{$P/a$}} & anal. & comp. & error & anal. & comp. & error
\\\hline
 0      &  2.00000 & 1.97825 & 1.09 & 2.00000 & 1.96345 & 1.83\\ 
 5      &  5.38516 & 5.34288 & 0.78 & 2.71750 & 2.66409 & 1.96 \\ 
 10     & 10.1980 & 10.0669 & 1.28 & 3.49257 & 3.42347 & 1.98 \\ 
 17.5   & 17.6139 & 17.2142 & 2.27 & 4.42876 & 4.33908 & 2.02 \\ 
\end{tabular}
\caption{Computed versus analytic values of the ADM energy and mass
for different values of $P$.}
\label{table:one}
\end{table}

We also computed the the pointwise relative error of $\psi$ averaged over the
vertices to compare with values reported by Cook \cite{Cook}.  For $P/a=10$, for instance,
we obtained an average pointwise relative error of 0.30\% using 59,248 vertices
and solving as a three-dimensional problem.  Cook reported an error
of 0.17\% solving as two-dimensional problem using a finite difference
method on a grid with 393,216 points (a three-dimensional grid with similar
mesh spacing would require about 250,000,000 points).

In addition to the radial problem just mentioned, Bowen and York proposed
values of $H$ which lead to two-dimensional initial data problems for
a single black hole.  Our computational results for these problems
were qualitatively very similar to those reported for the radial case.
They can be found in \cite{Mukherjee}.

\begin{figure}[!tbhp]
\centering
\includegraphics[width=3.5in,angle=270]{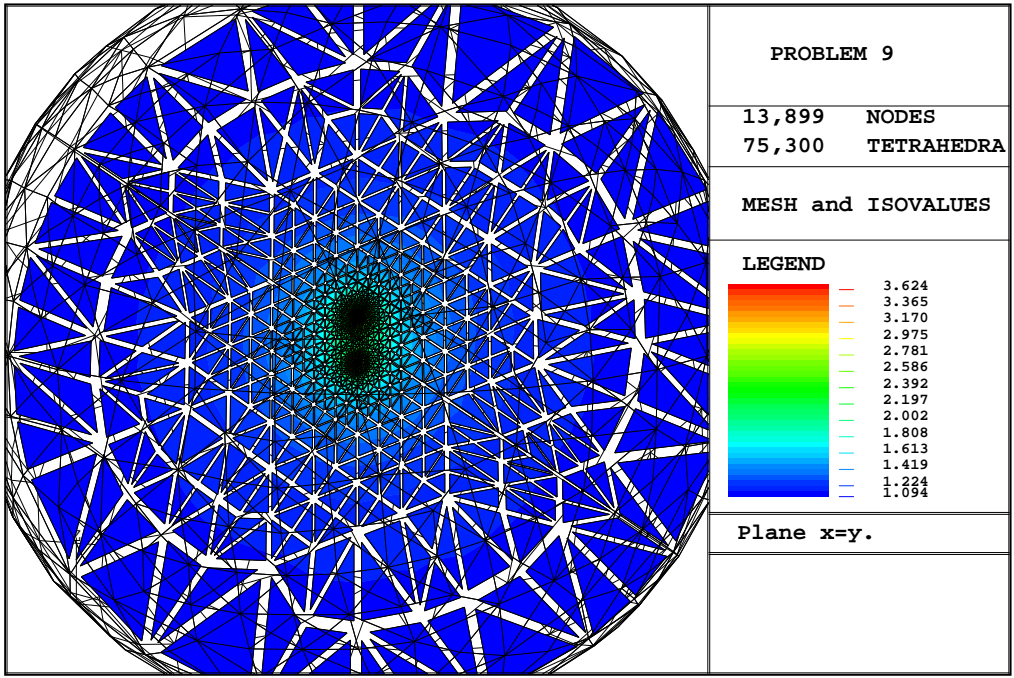}
\caption{Solution to a two black hole problem.}
\label{fig:bh1}
\end{figure}

\begin{figure}[!tbhp]
\centering
\includegraphics[width=3.5in,angle=270]{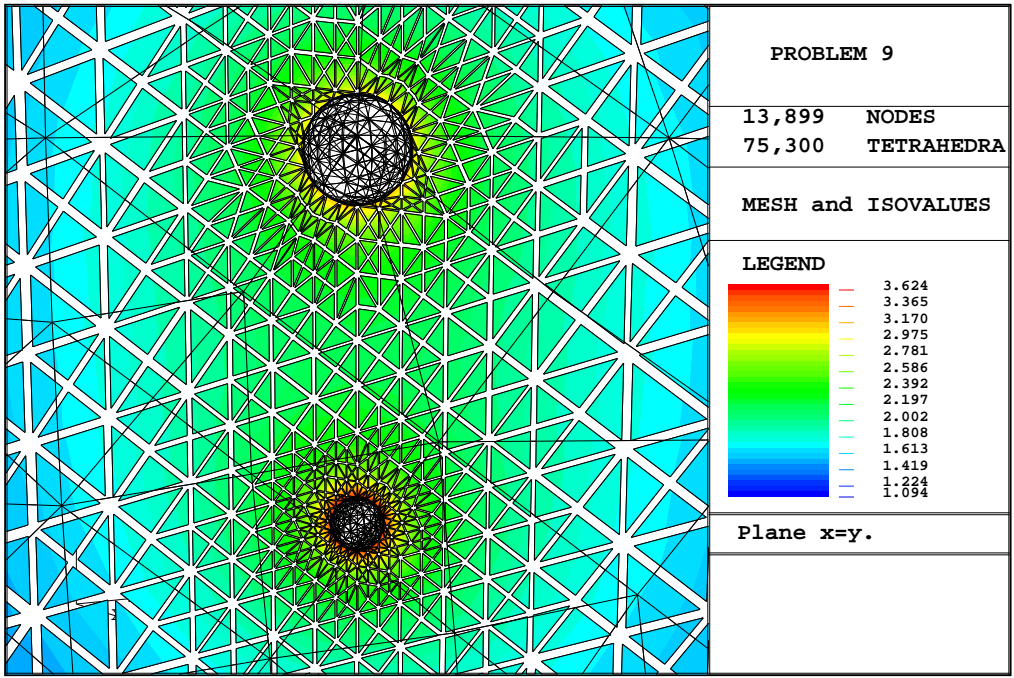}
\caption{Solution to a two black hole problem, zoom.}
\label{fig:bh2}
\end{figure}

\begin{figure}[!tbhp]
\centering
\includegraphics[width=3in]{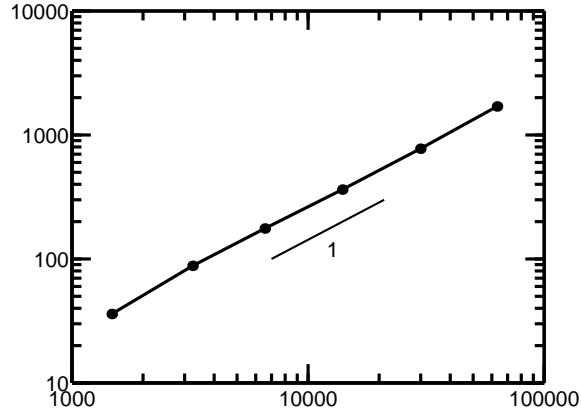}
\caption{CPU seconds, on the $y$-axis, versus number of degrees
of freedom, on the $x$-axis, for a binary black hole problem.}
\label{fig:times}
\end{figure}

Finally we describe the results of a binary black hole initial data
computation.  The black holes have radii $a=\sqrt3/2$ and $2a$ and their
centers are at $(0,0,-b)$ and $(0,0,b)$ respectively, where $b=2\sqrt3$.
The large ball has radius $128a$ and center at the origin.  The holes
are given linear momenta of $(0,0,15)$ and $(0,0,-15)$ respectively,
and no angular momenta.  The coarsest mesh had 585 vertices and 2,892
tetrahedra, while the finest mesh had 63,133 vertices and 346,084
tetrahedra.  In Figures~\ref{fig:bh1} and \ref{fig:bh2}, which is a zoom
of the previous figure, we show results computed on an intermediate
mesh with 13,899 vertices and 75,300 tetrahedra.  The figures show a
contour plot of $\psi$ on the plane $x=y$ (the plot shade is keyed
to the value of $\psi$).  This is easier to interpret in the color
version, which can be found in \cite{Mukherjee}.  The intersections
of the tetrahedra with the plane are shown slightly shrunk to improve
visibility.  Also shown are the mesh edges which intersect the boundary.
Finally, figure~\ref{fig:times} shows a plot of CPU time on a 1993 DEC
3000 model 500 with a single 150 MHz Alpha processor) versus number of
mesh vertices.  The plot clearly shows that the computation time is very
nearly proportional to the number of degrees of freedom.

\acknowledgement{The work of the first author was supported by NSF grant
DMS-9500672.  The work of the third author was supported by the Swiss
National Research Foundation.}
\bigskip

\noindent
Douglas N. Arnold\\
Department of Mathematics\\
Penn State University\\
University Park, PA   16802\\
USA
\medskip\\
Arup Mukherjee\\
Department of Mathematics\\
Rutgers University\\
New Brunswick, NJ  08903\\
USA
\medskip\\
Luc Pouly\\
ELCA Informatique SA\\
CH-1000 Lausanne\\
Switzerland

\end{document}